\documentclass[prd,superscriptaddress,twocolumn,%
  showpacs,amsmath,amssymb]{revtex4-1}
\usepackage{graphicx}
\usepackage[breaklinks, colorlinks=true]{hyperref}
\usepackage{color}

\renewcommand{\Re}{\mathrm{Re}\,}
\renewcommand{\Im}{\mathrm{Im}\,}

\begin{document}

\title{Analytical studies of the complex Langevin equation with
  a Gaussian Ansatz and multiple solutions in the unstable region}
\author{Yuya Abe}
\author{Kenji Fukushima}
\affiliation{Department of Physics, The University of Tokyo, %
  7-3-1 Hongo, Bunkyo-ku, Tokyo 113-0033, Japan}

\begin{abstract}
  We investigate a simple model using the numerical simulation in the
  complex Langevin equation (CLE) and the analytical approximation
  with the Gaussian Ansatz.  We find that the Gaussian Ansatz captures
  the essential and even quantitative features of the CLE results
  quite well including unwanted behavior in the unstable region where
  the CLE converges to a wrong answer.  The Gaussian Ansatz is
  therefore useful for looking into this convergence problem and we
  find that the exact answer in the unstable region is nicely
  reproduced by another solution that is naively excluded from the
  stability condition.  We consider the Gaussian probability
  distributions corresponding to multiple solutions along the
  Lefschetz thimble to discuss the stability and the locality.
  Our results suggest a prescription to improve the convergence of the
  CLE simulation to the exact answer.
\end{abstract}
\maketitle

%%%%%%%%%%   Introduction   %%%%%%%%%%
\section{Introduction}

Functional integration in quantum field theories shares the same
theoretical structure with the partition function in statistical
mechanics and the Monte-Carlo algorithm is the most useful for both
cases as long as the integrand is positive semi-definite.  However,
such an algorithm based on importance sampling in general breaks down
for the integrand having an oscillating complex phase.  This is the
notorious sign problem and it unfortunately appears in many
interesting physical environments such as quantum chromodynamics (QCD)
at finite baryon chemical potential, QCD with a $\theta$-term, Hubbard
model away from the half filling, imbalanced or frustrated spin
systems, etc.  As found in reviews on the sign
problem~\cite{LGSWSS90,Muroya:2003qs,Ejiri:2008nv,Fukushima:2010bq,%
deForcrand:2010ys,Aarts:2013lcm},
many ideas have been proposed and tested, but no solution is
established yet.  For finite-density QCD, introductory lectures are
available at Ref.~\cite{Aarts:2015tyj}.

It would be a natural idea to seek for an alternative quantization
scheme that is suitable for numerical simulations and does not rely on
importance sample.  One of the most promising candidates is the
so-called stochastic quantization using the Langevin equation or the
Fokker-Planck equation.  A comprehensive review is found in
Ref.~\cite{Damgaard:1987rr}.  An extension of the stochastic
quantization procedure to a theory with complex terms is specifically
referred to as the method of the complex Langevin equation (CLE), for
the Langevin variables are complexified then (for the early pioneering
attempt, see Ref.~\cite{Karsch:1985cb}).  The potential of the CLE has
been recently revisited as a theoretical tool to evade the sign
problem in finite-density QCD (see Ref.~\cite{Aarts:2013uxa} for a
modern review).  It was also expected that the CLE approach would be
capable of describing real-time dynamics~\cite{Berges:2005yt}, but it
was reported that the long-time numerical simulation falls into a
wrong solution~\cite{Berges:2005yt,Anzaki:2014hba} (see also
Ref.~\cite{Fukushima:2014iqa} for another real-time subtlety to define
the retarded and the advanced propagators with the CLE.)  To identify
the subtleties in the convergence problem in the CLE method, toy
models have been quite useful to provide us with insights about the
validity, which include low-dimensional
models~\cite{Pawlowski:2013pje,Nagata:2015uga}, (chiral) matrix
models~\cite{Nagata:2015uga,Mollgaard:2013qra}, and also even simple
1-dimensional (or called 0-dimensional in the field-theory context)
integrals~\cite{Nagata:2015uga,Aarts:2010gr,Aarts:2012ft,Aarts:2013uza,%
Nishimura:2015pba,Hayata:2015lzj}, some of which are motivated by the
sign problem for the Bose gas at finite chemical
potential~\cite{Aarts:2008wh,Aarts:2009hn}.

The breakthrough that triggered successful QCD (or gauge theory more
generally) simulations such as pioneering
Refs.~\cite{Sexty:2013ica,Aarts:2014bwa} and more recent
Refs.~\cite{Fodor:2015doa,Aarts:2016qrv} was the
recognition of the technique called the gauge cooling, which was
introduced to make the probability distribution not spread in the
complexified direction~\cite{Seiler:2012wz} (see also
Ref.~\cite{Aarts:2008rr} for successful U(1) and SU(3) link model
studies before the invention of the gauge cooling machinery), which
may also cure the problem caused by the drift term
singularity~\cite{Nagata:2015uga}.  On the formal level the
(sufficient) convergence criteria to the correct physical
answer are known~\cite{Aarts:2013uza,Aarts:2009uq}, which requires
analyticity (holomorphicity) of the theory and the locality of the
probability distribution.

It is sometimes quite instructive to consider the CLE method from the
point of view of a similar complexified approach known as the
Lefschetz thimble method as discussed in
Refs.~\cite{Hayata:2015lzj,Aarts:2013fpa,Aarts:2014nxa,%
Fukushima:2015qza,Tsutsui:2015tua}.
The Lefschetz thimble is a higher-dimensional extension of the
steepest descent path in complex analysis and the most important
property is that the complex phase is constant along this path or
thimble (see Refs.~\cite{Witten:2010cx} for mathematical foundation
and also Refs.~\cite{Cristoforetti:2012uv,Scorzato:2015qts} for recent
reviews).  The method has been implemented for quantum field
theory~\cite{Fujii:2013sra,Cristoforetti:2012su} and tested in
low-dimensional models~\cite{Tanizaki:2014tua,Fujii:2015bua} (see also
Ref.~\cite{Mukherjee:2014hsa} for a condensed matter application).
The important insight obtained from the Lefschetz thimble method is
that the Stokes phenomenon makes the structure of the theory
complicated~\cite{Kanazawa:2014qma}, which is the case near the phase
transition or in the real-time formalism~\cite{Fukushima:2015qza}.
Another interesting observation is that there may appear multiple
saddle-points that have complex phases and their destructive
interference~\cite{Hayata:2015lzj,Tanizaki:2015rda} is indispensable
to understand some non-trivial phenomenon like the Silver Blaze
puzzle~\cite{Cohen:2003kd,Splittorff:2006fu}.  The strong advantage in
the Lefschetz thimble method is that the analytical investigations are
possible, which also leads to a new discovery of hidden theoretical
structures~\cite{Behtash:2015kna}.

The objective of this work is to explore some analytical aspects in
the CLE approach.  As compared to the Lefschetz thimble especially for
the 1-dimensional integral models, the CLE studies more often rely on
numerical simulations.  As closely discussed in
Ref.~\cite{Aarts:2013uza} the probability distribution function can be
constructed perturbatively, but to reveal the full profile, the
numerical calculations are unavoidable, which is of course useful to
deepen our understanding, but it would be desirable if we have
analytical formulas from which we can somehow infer detailed
information on the theory.  To this end, we would propose a Gaussian
Ansatz in the present paper.  This is a generalization of the
mean-field treatment to the CLE framework.  The idea can be traced
back to a variational approach to stochastic
quantization~\cite{Amundsen:1983tg}, and it was reported that an
analytical evaluation with one variational parameter (corresponding to
the dynamical mass) agrees quite well with the full numerical result
for a 1-dimensional $N$-component model.  Within this variational
approach the $1/N$ expansion has been also discussed in
Ref.~\cite{Grandati:1992hj}.  Such a mean-field treatment has been
generalized to the CLE with a complex action for the relativistic Bose
gas at finite chemical potential~\cite{Aarts:2009hn}.  This direction
of extension should be quite intriguing;  for example, a comparison
between the mean-field results and the CLE results in the Polyakov
loop model has provided us with a useful hint on the breakdown of the
CLE with a branch-cut crossing problem~\cite{Greensite:2014cxa}, and
the more direct mean-field treatment of the CLE method itself would
give us a further analytical insight into the subtlety of the
convergence, as we will discuss.  (For the Lefschetz thimble version
of the comparison to the mean-field Polyakov loop model, see
Ref.~\cite{Tanizaki:2015pua} that has justified the mean-field
treatment in Ref.~\cite{Fukushima:2006uv}).

This paper is organized as follows.  In Sec.~\ref{sec:formalism} basic
equations of the CLE method are summarized for convenience of readers
and the Gaussian Ansatz is introduced with two variational parameters
as a generalization of the free two-point function.
Section~\ref{sec:model} is devoted to detailed explanations of the
properties of the 1-dimensional quartic model, followed by the main
part of this paper in Sec.~\ref{sec:analysis} in which a comparison
between the Gaussian Ansatz results and the exact answer is made for
three distinct regions of the model parameters.  Conclusion is finally
given in Sec.~\ref{sec:conclusion}.

%%%%%%%%%%   Formalism   %%%%%%%%%%
\section{Formalism}
\label{sec:formalism}

We briefly look over the general formalism of the complex Langevin
method and its equivalent representation using the Fokker-Planck
equation.  Then, we introduce our idea of the Gaussian Ansatz as an
approximate solution of the Fokker-Planck equation.  Here we will
present expressions for a scalar field theory only, but the
generalization for other field theories should be straightforward.

The fundamental ingredient in the complex Langevin method is a
complexified extension of the Langevin equation with a fictitious time
$\tau$, which reads,
\begin{equation}
  \frac{\partial \phi(x,\tau)}{\partial \tau} =
  -\frac{\delta S[\phi]}{\delta \phi(x,\tau)} + \eta(x,\tau)\;,
  \label{eq:CL}
\end{equation}
where $\eta(x,\tau)$ represents stochastic noise satisfying
$\langle\eta(x,\tau)\eta(x',\tau')\rangle
=2\delta^{(d)}(x-x')\delta(\tau-\tau')$.  If the action $S$ takes
a complex value, as is the case for fermions with a finite chemical
potential or general real-time dynamics, $\phi$ should be also
complexified as $\phi = \phi_R + i\phi_I$ with
$\phi_R, \phi_I\in \mathbb{R}$.  For analytical purposes it is often
more convenient to deal with a different but equivalent representation
of the quantization procedure using the Fokker Planck equation, that
is expressed as
\begin{equation}
  \begin{split}
    \frac{d P[\phi]}{d\tau} &= \int d^d x\,\biggl\{
    \frac{\delta}{\delta\phi_R} \biggl[ \Re\biggl(
      \frac{\delta S}{\delta\phi} \biggr) P[\phi] \biggr] \\
    &\qquad\qquad + \frac{\delta^2 P}{\delta\phi_R^2}
    + \frac{\delta}{\delta\phi_I}\biggl[ \Im\biggl(
      \frac{\delta S}{\delta\phi}\biggr) P[\phi] \biggr]\biggr\}\;.
  \end{split}
  \label{eq:FPeq}
\end{equation}
For sign-problem free field theories in Euclidean space-time, the
action $S$ is a real functional of real $\phi$ and the solution of
Eq.~\eqref{eq:FPeq} approaches $P[\phi]\propto e^{-S[\phi]}$.  In
Minkowskian space-time, on the other hand, the action is complex (and
another simple but useful example of a complex action is the Bose gas
at finite chemical potential~\cite{Aarts:2008wh,Aarts:2009hn}).  In
a free real-time scalar theory, for an explicit example, $S$ in
momentum space is complex as
\begin{equation}
  S[\phi] = i\int\frac{d^d p}{(2\pi)^d}\,
  \phi(-p)(-p^2 + m^2-i\epsilon)\phi(p)\;,
  \label{eq:Smin}
\end{equation}
where a small real part is necessary for convergence in the
$i\epsilon$ prescription.  Obviously a real valued $P[\phi]$ cannot
approach a standard form of the functional integral weight
$\sim e^{-S[\phi]}$ because the weight is complex then.  It is quite
instructive that an analytical solution of the Fokker-Planck
equation~\eqref{eq:FPeq} is known for this example of
Eq.~\eqref{eq:Smin} as~\cite{Damgaard:1987rr}
\begin{align}
  P[\phi] = N\exp\biggl\{ &-\int\frac{d^d p}{(2\pi)^d}\,\epsilon
  \biggl[\phi_R(-p)\phi_R(p) \notag\\
    & + \biggl(1+\frac{2\epsilon^2}{(p^2-m^2)^2}\biggr)
    \phi_I(-p)\phi_I(p) \notag\\
    & - \frac{2\epsilon}{p^2-m^2}\phi_R(-p)\phi_I(p)\
    \biggr]\biggr\}\;.
  \label{eq:Gaussian}
\end{align}
It is just a straightforward calculation to confirm that we can
recover a correct expression for the propagator from this real
probability weight, i.e.
\begin{equation}
  \int \mathcal{D}\phi_R\,\mathcal{D}\phi_I\, P[\phi]\,\phi(-p)\phi(p)
  = \frac{i}{p^2-m^2+i\epsilon}\,.
  \label{eq:prop}
\end{equation}
We should note that the imaginary part in the right-hand side in
Eq.~\eqref{eq:prop} arises from not the weight $P[\phi]$ but
complexified $\phi(p)$ in the left-hand side.  Therefore, in such
complexified representation of theory, the sign problem is evaded but
the operator generally acquires residual complex phase.

The prescription we would propose in this work is a Gaussian Ansatz as
an extension of Eq.~\eqref{eq:Gaussian}, namely,
\begin{equation}
  \epsilon \to A(p)\;,\qquad p^2-m^2 \to -B(p)
  \label{eq:Ansatz}
\end{equation}
for interacting field theories.  Here, real-valued $A$ and $B$ are
to be regarded as ``renormalized'' width and mass including
interaction effects and should be determined by the stationary
condition of the Fokker-Planck equation, that is; $dP/d\tau=0$.
Conceptually, the above Ansatz should correspond to a Gaussian
truncation with ``mean-field'' variables $A$ and $B$ optimized by the
variational principle.  Although this Ansatz introduces an
approximation, a fully analytical treatment is feasible then and it
should be useful to understand how the complex Langevin equation
converges to a false solution (which has been understood from
power-decay behavior of the probability
distribution~\cite{Aarts:2013uza} but we will shed light from a
different perspective) and where we can find a correct answer (for
successful applications of the mean-field treatment, see
Ref.~\cite{Aarts:2009hn}).

%%%%%%%%%%   Quartic Model   %%%%%%%%%%
\section{Quartic Model}
\label{sec:model}

Here, a simplest 1-dimensional (or in the field-theory context, it is
commonly called ``0-dimensional'' counting the number of spacetime)
example should suffice for our present purpose to demonstrate how
useful the Gaussian Ansatz is to get an analytical insight.

%%%%%   Definition   %%%%%
\subsection{Definition}

We define the ``theory'' by the following integral~\cite{KP1984};
\begin{equation}
  \begin{split}
    & Z(\alpha,\beta) = \int d\phi\, e^{-S(\phi;\alpha,\beta)}\;,\\
    & S(\phi;\alpha,\beta) = \frac{1}{2}\alpha \phi^2
    +\frac{1}{4}\beta \phi^4\;,
  \end{split}
  \label{eq:model}
\end{equation}
where $\alpha = a+ib$ ($a,b\in\mathbb{R}$) and $\beta=c+id$
($c,d\in\mathbb{R}$) are complex coefficients and $\phi$ is a real
integration variable.  After the $\phi$ integration we can find the
exact result in terms of the modified Bessel function as
\begin{equation}
  Z(\alpha,\beta) = \sqrt{\frac{\alpha}{2\beta}}\;
  e^{\alpha^2/(8\beta)} K_{1/4}\biggl(\frac{\alpha^2}{8\beta}\biggr)
  \label{eq:exact}
\end{equation}
for $\Re\alpha>0$ and $\Re\beta>0$.  For $\Re\alpha<0$ the Bessel
function in the above expression should be replaced with
$I_{\pm1/4}\bigl(\frac{\alpha^2}{8\beta}\bigr)$.  For the validity
check of the method, we will refer to the \textit{exact answer} that
we can obtain from these analytical expressions or from the direct
numerical integration of Eq.~\eqref{eq:model}.

This theory has interesting features similar to phase structures.  To
see them let us consider a two-point function, that is,
\begin{align}
  \langle\phi^2\rangle_{\text{exact}} &=
  \frac{\int d\phi\,\phi^2 e^{-S}}{\int d\phi\, e^{-S}} \notag\\
  &= \frac{\alpha}{4\beta}\frac{K_{-3/4}\bigl(\frac{\alpha^2}
    {8\beta}\bigr) + K_{5/4}\bigl(\frac{\alpha^2}{8\beta}\bigr)}
       {K_{1/4}\bigl(\frac{\alpha^2}{8\beta}\bigr)} - \frac{\alpha}
       {2\beta} - \frac{1}{\alpha}
\end{align}
for $\Re\alpha>0$ and $\Re\beta>0$.  Again, it is not difficult to
carry out the direct numerical integration as long as $\Re\beta>0$.
Now we see $\langle\phi^2\rangle_{\text{exact}}$ as a function of
$a=\Re\alpha$ and $b=\Im\alpha$ while keeping $\beta=1$.  Such a
choice does not loose the generality because we can always rescale
$\phi$ (after complexifying the theory) so that $\beta=1$.

%---   figure   ---%
\begin{figure}
  \includegraphics[width=\columnwidth]{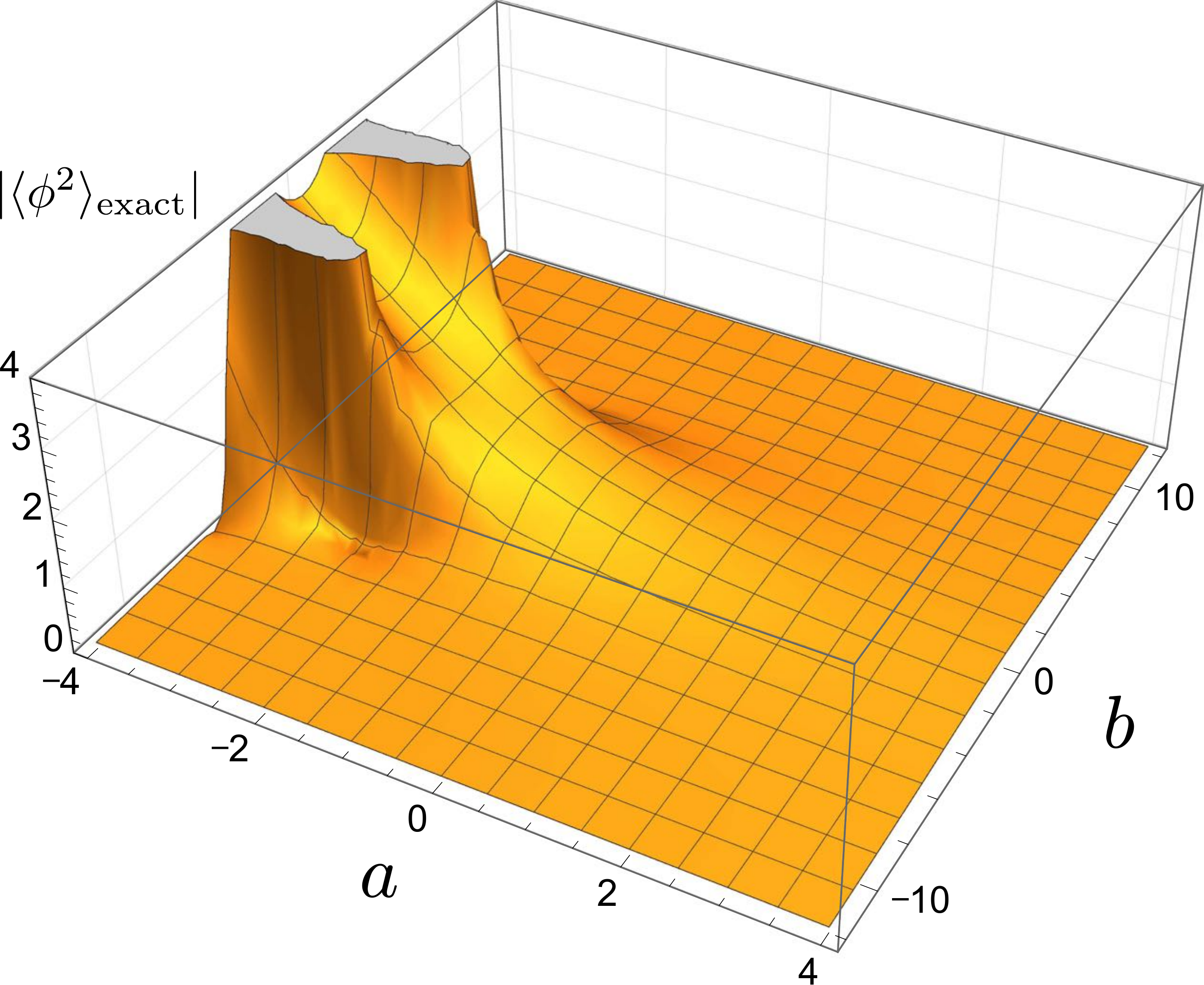}
  \caption{Absolute value of the exact answer of
    $\langle\phi^2\rangle_{\text{exact}}$ as a function of the real
    part $a$ and the imaginary part $b$ of the quadratic coefficient
    $\alpha$.}
 \label{fig:exact}
\end{figure}
%---   figure   ---$

Figure~\ref{fig:exact} shows $|\langle\phi^2\rangle_{\text{exact}}|$
as a function of $a$ and $b$.  It is clear to see that the expectation
value increases in the region for $a < 0$, which is reminiscent of the
spontaneous symmetry breaking.  Of course, the present model does not
have infinite degrees of freedom and, strictly speaking, the
spontaneous symmetry breaking is impossible,
i.e.\ $\langle\phi\rangle_{\text{exact}}=0$ always holds for any $a$.
Nevertheless, we can understand that a situation similar to the
spontaneous symmetry breaking occurs in the following sense.  The
integration~\eqref{eq:model} is dominated around the minima
(saddle-points) of $S$ as obtained from $dS/d\phi=0$.  For the present
theory there are three saddle-points,
\begin{equation}
  \bar{\phi}_0 = 0\;,\qquad
  \bar{\phi}_\pm = \pm \sqrt{-\alpha/\beta}\;.
\end{equation}
Clearly $S=S_0=0$ at $\phi=\bar{\phi}_0$ and $S=S_\pm=-\alpha^2/4$ at
$\phi=\bar{\phi}_\pm$.  Therefore, as long as $\Re S_\pm < \Re S_0$,
the integration is dominated around $\bar{\phi}_\pm$ and
$\langle\phi^2\rangle_{\text{exact}}$ should behave like
$-\alpha/\beta$ in the first approximation.  There is no phase
transition in a strict sense, but we may well identify this situation
physically as an analogue of the \textit{ordered state}.

The qualitative behavior, however, changes drastically at
$|b|\simeq |a|$ with $a<0$ and, so to speak, the broken symmetry is
restored in the region for $|b|>|a|$.  The reason for this change is
easy to understand from the above consideration.  For $|b|>|a|$ we
explicitly see $\Re S_\pm=-a^2+b^2 > \Re S_0 = 0$, and so the
integration is again dominated around $\bar{\phi}_0$ and the non-zero
expectation values at $\bar{\phi}_\pm$ become irrelevant in effect.
Thus, we may say that the ordered state is hindered by attenuation
effects caused by large $b$.

In summary this theory has three characteristic and qualitatively
distinct states depending on $a$ and $b$ as follows:
\begin{center}
  \begin{tabular}{l p{0.6em} l}
    $a > 0$\;: && Normal State \\
    $a < 0,\;\; a^2 \gtrsim b^2$\;:  && Ordered State \\
    $a < 0,\;\; b^2 \gtrsim a^2$\;: && Attenuated State
  \end{tabular}
\end{center}
It is known that the CLE fails in the attenuated state for $a<0$ and
$b^2\gtrsim a^2$, which we will closely investigate analytically
using the Gaussian Ansatz.  Also, it would be worth while mentioning
that the convergence of the CLE simulation to the exact answer is
proven in the region with $a>0$ and $b^2<3a^2$, while the CLE
simulation may not work for higher order expectation values,
$\langle\phi^n\rangle$ $(n\ge 4)$, in the region with $a>0$ and
$b^2>3a^2$ even in the normal state~\cite{Aarts:2013uza}.  For the
moment we will focus on $\langle\phi^2\rangle$ and will test our
method for $\langle\phi^4\rangle$ later.

%%%%%   Results from the CLE   %%%%%
\subsection{Results from the CLE}

%---   figure   ---%
\begin{figure}
  \includegraphics[width=\columnwidth]{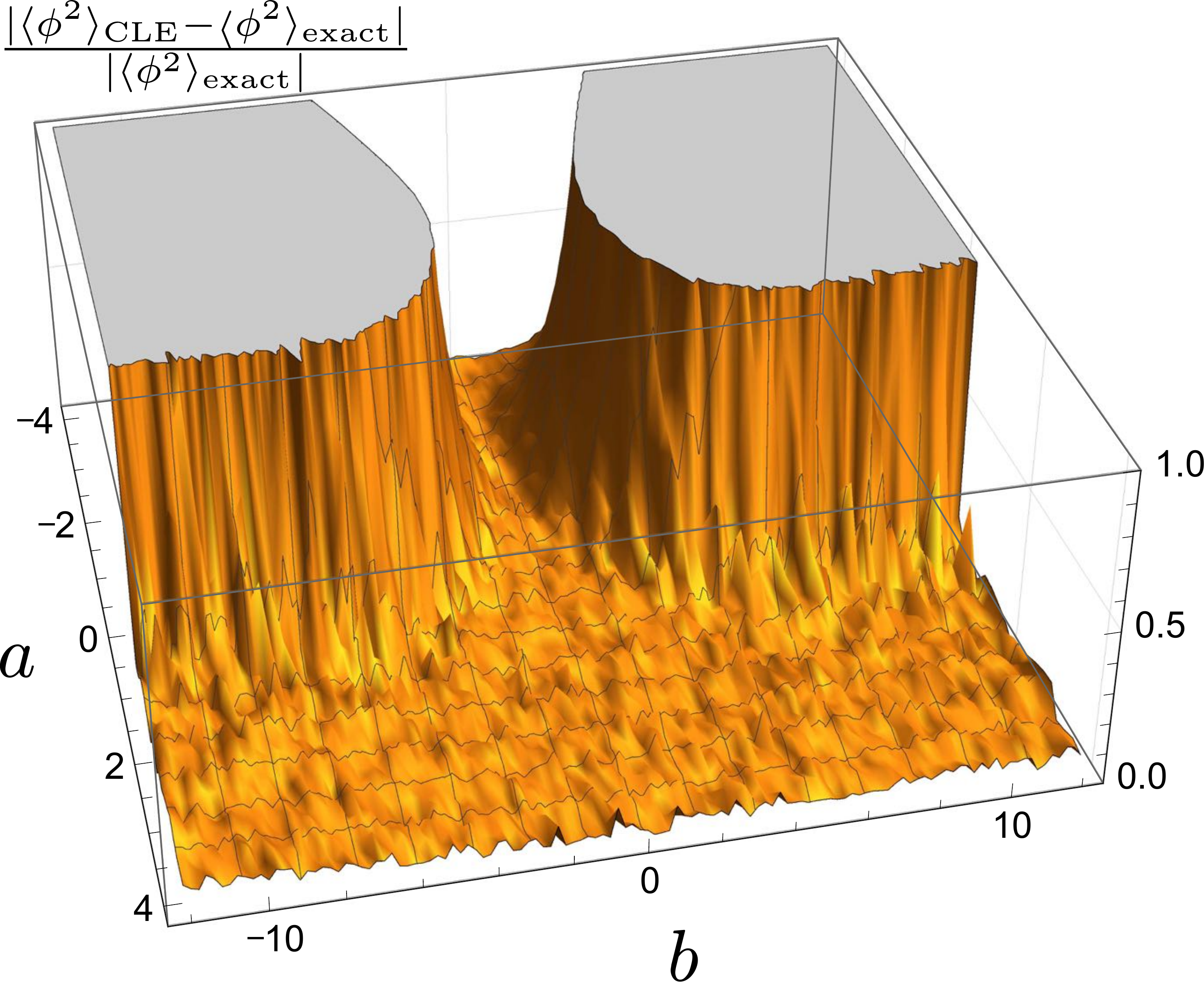}
  \caption{Comparison between the CLE results and the exact answer.}
 \label{fig:cle}
\end{figure}
%---   figure   ---$

Here, we briefly discuss the results from the CLE to show where the
CLE fails (for the two-point function).  We numerically solved
Eq.~\eqref{eq:CL} for the action~\eqref{eq:model} with discretization
$dt=5\times 10^{-3}$ and updated the fictitious time evolution by
$10^5$ steps and average the numerical outputs over time.  When we
detected a runaway trajectory, we took $10^3$ steps back to avoid
divergence.

We show the difference between the CLE results and the exact answer in
Fig.~\ref{fig:cle}.  From this comparison we see that the CLE works
good for $a>0$ generally (there may be a failure for the higher order
functions~\cite{Aarts:2013uza} and we will come to this point in the
end of this paper), and also it works in the ordered state with
$a<0$ as long as $a^2>b^2$.  We just note that in this theory the
Stokes phenomenon occurs at $a=0$, and so the onset of the Stokes
phenomenon does not necessarily coincide with the breakdown of the CLE
simulation.  As we mentioned before, the most problematic region for
the CLE calculation in this theory is $a<0$ and $b^2\gtrsim a^2$,
which we call the attenuated state throughout this present work.

%%%%%%%%%%   Analysis with the Gaussian Ansatz   %%%%%%%%%%
\section{Analysis with the Gaussian Ansatz}
\label{sec:analysis}

Because the Gaussian Ansatz is not an exact solution of the
Fokker-Planck equation~\eqref{eq:FPeq} for $\beta\neq 0$ in general,
there is some ambiguity in the determination of $A$ and $B$ associated
with the choice of what we optimize.  A prescription we adopt here is
an equilibrium condition for a two-point function, that is,
\begin{equation}
  \int d\phi_R\,d\phi_I\, \phi^2 \,\frac{d P[\phi]}{d\tau} = 0\;,
\label{eq:condition}
\end{equation}
which naturally must hold when $dP[\phi]/d\tau=0$ is reached.  From
the real and the imaginary parts of the above condition, we can get
two equations to solve $A$ and $B$ as functions of $a$ and $b$.  It
should be noted that this condition has similarity to the
``criteria for correctness'' of the second order as discussed in
Ref.~\cite{Aarts:2013uza}.  Thus our condition naturally leads to a
sort of gap equation in a sense that the $n$-th order ``criteria for
correctness'' is equivalent to the Schwinger-Dyson equation of
$n$-point function.

%%%%%   Normal State   %%%%%
\subsection{Normal State}

Plugging the action~\eqref{eq:model} into the Fokker-Planck
equation~\eqref{eq:FPeq}, we can express $dP/d\tau$ and substitute it
for Eq.~\eqref{eq:condition}.  We can explicitly perform the Gaussian
integrations with respect to $\phi_R$ and $\phi_I$ and after
simplifying terms, we can find a set of equations to fix
$A(\alpha,\beta)$ and $B(\alpha,\beta)$ as
\begin{align}
  A-a &= \frac{3}{A^2+B^2}(cA+dB) \;,\\
  B-b &= \frac{3}{A^2+B^2}(dA-cB) \;.
\label{eq:MFeq}
\end{align}
It is easy to confirm that the non-interacting limit at $\beta=0$
(i.e.\ $c=d=0$) immediately leads to $A(\alpha,0)=a$ and
$B(\alpha,0)=b$ as it should (and this free solution is nothing but
the lowest order solution discussed in Ref.~\cite{Aarts:2013uza}).

There are four independent branches of solutions for
Eq.~\eqref{eq:MFeq}.  For our present choice of $\beta=1$ (i.e.\ $c=1$
and $d=0$) the above set of equations leads to two complex and
two real solutions.  The real solution with $A>0$, which is required
for the stability of the Gaussian integration, is uniquely determined
as
\begin{align}
  A &= \frac{a}{2} + \frac{1}{2\sqrt{2}}
  \sqrt{\!\sqrt{48a^2+(a^2\!+\!b^2\!-\!12)^2} + a^2\!-\!b^2\!+\!12}\;,
  \label{eq:MFA}\\
  B &= \frac{b}{2} + \frac{\text{sgn}(ab)}{2\sqrt{2}}
  \sqrt{\!\sqrt{48a^2+(a^2\!+\!b^2\!-\!12)^2} - a^2\!+\!b^2\!-\!12}\;.
  \label{eq:MFB}
\end{align}
It is clear that we can give an estimate for the two-point function
using the Gaussian Ansatz in the following way,
\begin{equation}
  \langle\phi^2\rangle_{\text{Gauss}} = \frac{\int d\phi_R\,d\phi_I\,
    \phi^2 P[\phi]}{\int d\phi_R\,d\phi_I\, P[\phi]}
  = \frac{1}{A(\alpha,\beta)+iB(\alpha,\beta)}\;.
\label{eq:MF-GF}
\end{equation}
This final result from the Gaussian Ansatz is so simple in the
analytical structure as compared to the exact answer, but we can
confirm that this gives a good approximation in the normal state.

%---   figure   ---%
\begin{figure}
  \includegraphics[width=\columnwidth]{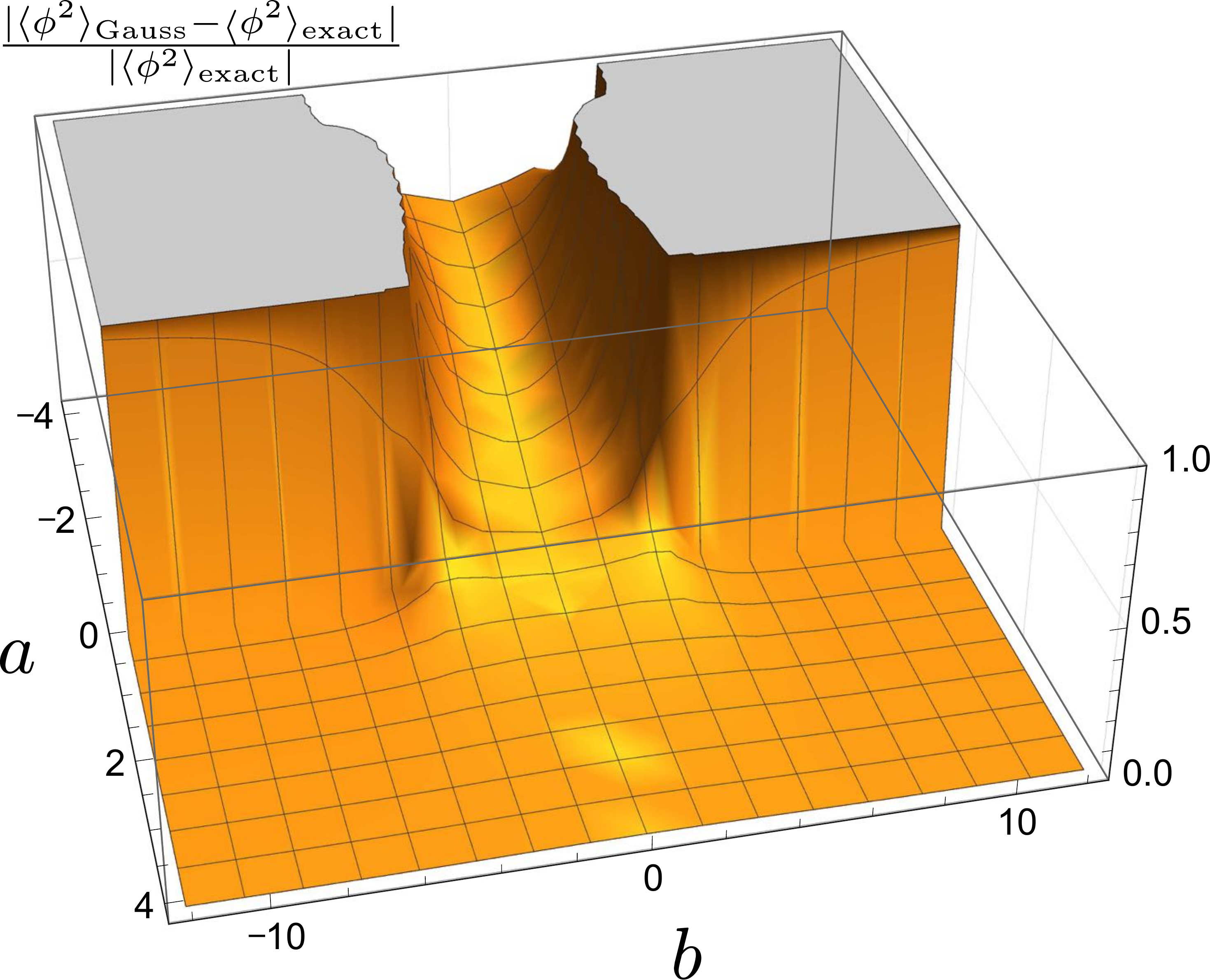}
  \caption{Comparison between the Gaussian Ansatz results and the
    exact answer for the quadratic operator.}
  \label{fig:gaussNS}
\end{figure}
%---   figure   ---$

To quantify how good the Gaussian Ansatz is, let us make a plot in a
way similar to Fig.~\ref{fig:cle}, with
$\langle\phi^2\rangle_{\text{CLE}}$ replaced with
$\langle\phi^2\rangle_{\text{Gauss}}$, which is presented in
Fig.~\ref{fig:gaussNS}.  It is very interesting to see that
Fig.~\ref{fig:gaussNS} has a remarkable similarity to
Fig.~\ref{fig:cle}.  In particular the Gaussian Ansatz works
excellently to show good agreement with the exact answer in the region
with $a\ge 0$ for any $b$.

Our Gaussian Ansatz takes care of fluctuations around
$\phi=\bar{\phi}_0$ and implicitly neglects the contributions from
other saddle-points at $\phi=\bar{\phi}_\pm$.  In terms of the
Lefschetz thimble method, such an approximated treatment is justified
by the fact that only the thimble attached to $\bar{\phi}_0$ makes a
finite contribution for $a>0$.  There is, however, a sudden change in
the thimble structure at $a=0$, which is commonly called the Stokes
phenomenon, and eventually all three thimbles attached to
$\bar{\phi}_0$ and $\bar{\phi}_\pm$ come to make a finite contribution
for $a<0$.  This sudden change partially explains the sudden breakdown
of the Gaussian Ansatz estimate around $a=0$.

One might have an impression that the Gaussian Ansatz may still work
in the ordered state in view of Fig.~\ref{fig:gaussNS} but some
cautions are needed.  In the region with $a^2>b^2$ and $a<0$, as $|a|$
grows up (and so the ``condensate'' grows up), the agreement gets
worse.  For example, at $a=-4$ and $b=0$ as shown in
Fig.~\ref{fig:gaussNS},
$|\langle\phi^2\rangle_{\text{Gauss}}-\langle\phi^2\rangle_{\text{exact}}|
/|\langle\phi^2\rangle_{\text{exact}}|\approx 0.5$ and the deviations
would be larger with increasing $|a|$ in the negative direction.
However, as we see from Fig.~\ref{fig:cle}, the CLE should describe
the physics correctly also in this region of the ordered state.

%%%%%   Ordered State   %%%%%
\subsection{Ordered State}

In many physical examples the Stokes phenomenon makes the difficulty
even more difficult.  However, the CLE is capable of going beyond the
Stokes phenomenon from the normal state to the ordered state except
for the onset region.  Although the most interesting question is what
should be happening in the attenuated state as we will address later,
let us clarify how the Gaussian Ansatz can capture the correct physics
in the ordered state too.

In this region the contributions around $\bar{\phi}_\pm$ should be
dominant, and so the Gaussian Ansatz must be formulated also around
$\bar{\phi}_\pm$.  In the mean-field type calculations it is a common
technique to consider fluctuations around a shifted vacuum that is
self-consistently determined by the energy minimization condition.
Therefore, the probability weight should be changed as
$P[\phi]\to P_\pm[\phi]=P[\phi+\bar{\phi}_\pm]$ with $A_\pm$ and
$B_\pm$.  Then, the two-point function should be approximated as
\begin{equation}
  \langle\phi^2\rangle_{\text{Gauss}}
  = \frac{1}{A_\pm + iB_\pm} - \frac{\alpha}{\beta}\;,
\end{equation}
where the last term represents the contribution by
$\bar{\phi}_\pm^2=-\alpha/\beta$.  The set of equations to fix $A_\pm$
and $B_\pm$ is slightly changed from Eq.~\eqref{eq:MFeq} by
$\bar{\phi}_\pm^2$ as
\begin{align}
  A_\pm + 2a &= \frac{3}{A_\pm^2+B_\pm^2}(cA_\pm+dB_\pm)\;,\\
  B_\pm + 2b &= \frac{3}{A_\pm^2+B_\pm^2}(dA_\pm-cB_\pm)\;.
\end{align}
We can solve the above easily to find the analytical expressions
again.

Here, let us make a remark on the convergence to the right solution.
Generally speaking, beyond the Stokes phenomenon, multiple
saddle-points take part in the integration;  all of $\bar{\phi}_0$ and
$\bar{\phi}_\pm$ in the present case.  In physical applications it is
often the case that one of them dominates the physics.  The
spontaneous symmetry breaking is such a phenomenon that can be
correctly described by one saddle-point property.  Because our
1-dimensional theory does not have such symmetry breaking, we have to
sum up both contributions from $\bar{\phi}_\pm$ to have
$\langle\phi\rangle=0$, but with infinite degrees of freedom only one
contribution is spontaneously chosen, and thus the CLE method should
work better then.  Some other special examples are known to be more
problematic.  The mixed phase associated with a first-order phase
transition should be one of the most typical examples.  In such a
situation contributions from different saddle-points are equally
important, and also missing of a relative complex phase may cause a
further problem of falling into a wrong answer~\cite{Hayata:2015lzj}.
Another famous example is the Silver Blaze problem for which relative
phases from infinite saddle-points make destructive
interference~\cite{Tanizaki:2015rda}.  For these problems, to
formulate the Gaussian Ansatz to work, we need to take a proper
superposition of $P_0[\phi]$ centered at $\bar{\phi}_0$ and
$P_\pm[\phi]$ centered at $\bar{\phi}_\pm$ with relative weights
including complex phase factors.

%%%%%   Attenuated State   %%%%%
\subsection{Attenuated State}

The most interesting and non-trivial question is why the CLE
simulation and also the Gaussian Ansatz do not work in the
attenuated state with $a<0$ and $b^2 \gtrsim a^2$.  This is beyond the
Stokes phenomenon, but because of the real weight,
$e^{-\Re S}=e^{-(b^2-a^2)/4}\ll 1$, we can safely neglect these
contributions from $\bar{\phi}_\pm$ for $b^2\gg a^2$ and the
integration should be well approximated by the fluctuations around
$\bar{\phi}_0$ only.  Thus, it is very likely that the CLE simulation
and the Gaussian Ansatz should be valid descriptions, but they fail in
practice.  Because the Gaussian Ansatz enables us to cope with the
problem with simple analytical formulas, we can relatively easily
identify the source of the problem.

%---   figure   ---%
\begin{figure}
  \includegraphics[width=\columnwidth]{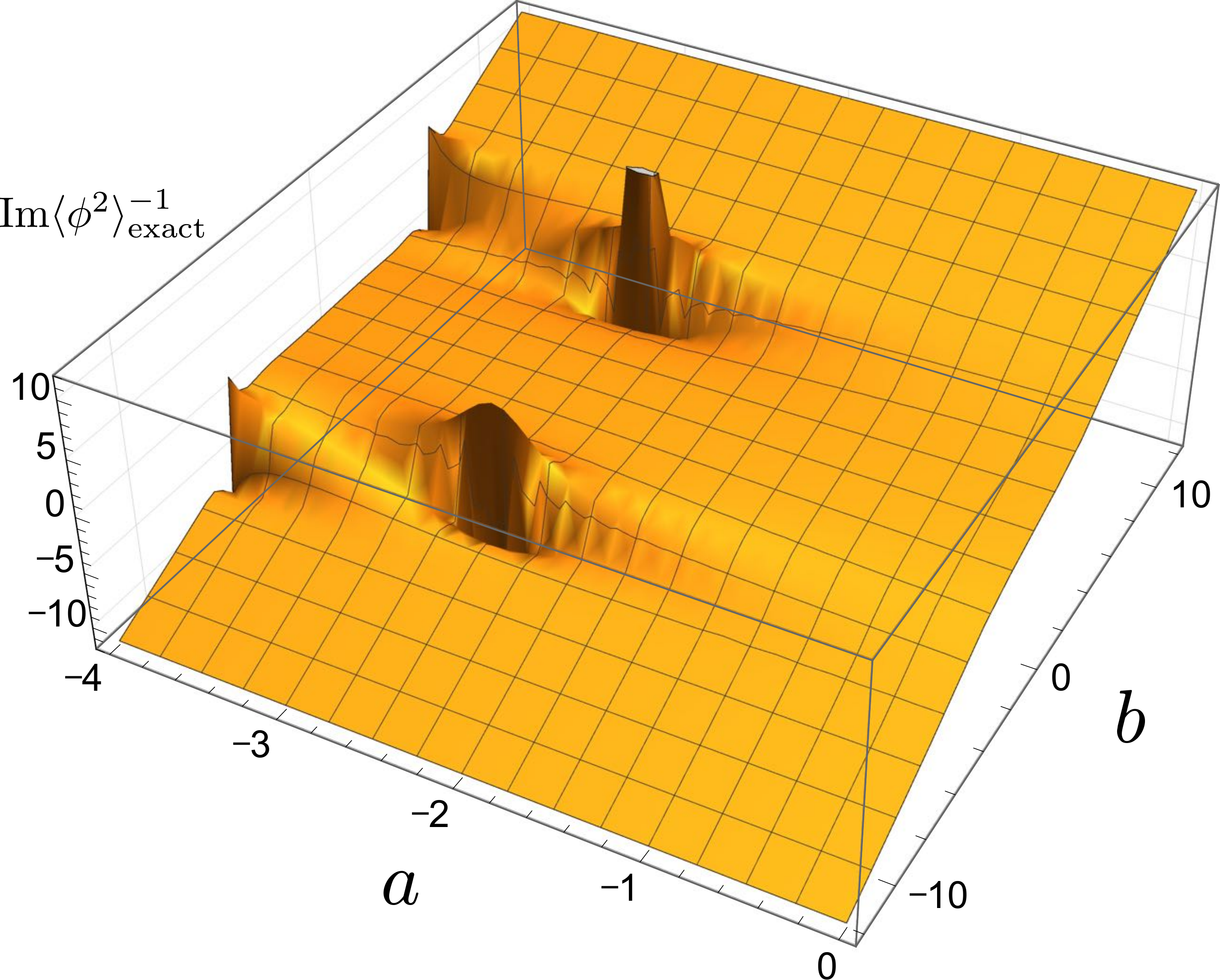}
  \caption{Counter part of $B$ inferred from
    $\langle\phi^2\rangle_{\text{exact}}$ as a function of $a$ and $b$
    in the region for $a<0$.}
  \label{fig:Bexact}
\end{figure}
%---   figure   ---$

It is already apparent from Eq.~\eqref{eq:MFB} how the Gaussian Ansatz
leads to unphysical behavior.  Let us consider expected behavior of
$B$ for asymptotically large $b$.  Naturally, we would immediately
anticipate $B\simeq b$ for large enough $b$ from our physical
intuition, and Eq.~\eqref{eq:MFB} indeed predicts $B\to b$ for
$|b|\gg a$ as long as $a>0$, while it gives $B\to 0$ for $|b|\gg |a|$
once the system enters $a<0$.  This is a very clear manifestation of
where the wrong answer is picked up.  Actually, the exact answer
certainly exhibits the behavior of $B\simeq b$ even in the $a<0$
region, as checked in Fig.~\ref{fig:Bexact} where a counter part of
$B$ defined by $\Im\langle\phi^2\rangle_{\text{exact}}^{-1}$ is
plotted as a function of $a$ and $b$.  Apart from the ordered state
for $a<0$, $a^2>b^2$ and some spiky structures near the phase border,
$\Im\langle\phi^2\rangle_{\text{exact}}^{-1}$ clearly scales as
$\sim b$ in a way consistent with our intuition. 

We already pointed out that there is another real solution of the set
of equations~\eqref{eq:MFeq}, whose explicit forms are
\begin{align}
  \tilde{A} &= \frac{a}{2} - \frac{1}{2\sqrt{2}}
  \sqrt{\!\sqrt{48a^2+(a^2\!+\!b^2\!-\!12)^2} + a^2\!-\!b^2\!+\!12}\;,
  \label{eq:MFA2}\\
  \tilde{B} &= \frac{b}{2} - \frac{\text{sgn}(ab)}{2\sqrt{2}}
  \sqrt{\!\sqrt{48a^2+(a^2\!+\!b^2\!-\!12)^2} - a^2\!+\!b^2\!-\!12}\;.
  \label{eq:MFB2}
\end{align}
The problem of this solution is that the integration with the Gaussian
probability is not well-defined due to $\tilde{A}<0$ and we should
usually exclude this branch of the solution.

%---   figure   ---%
\begin{figure}
  \includegraphics[width=\columnwidth]{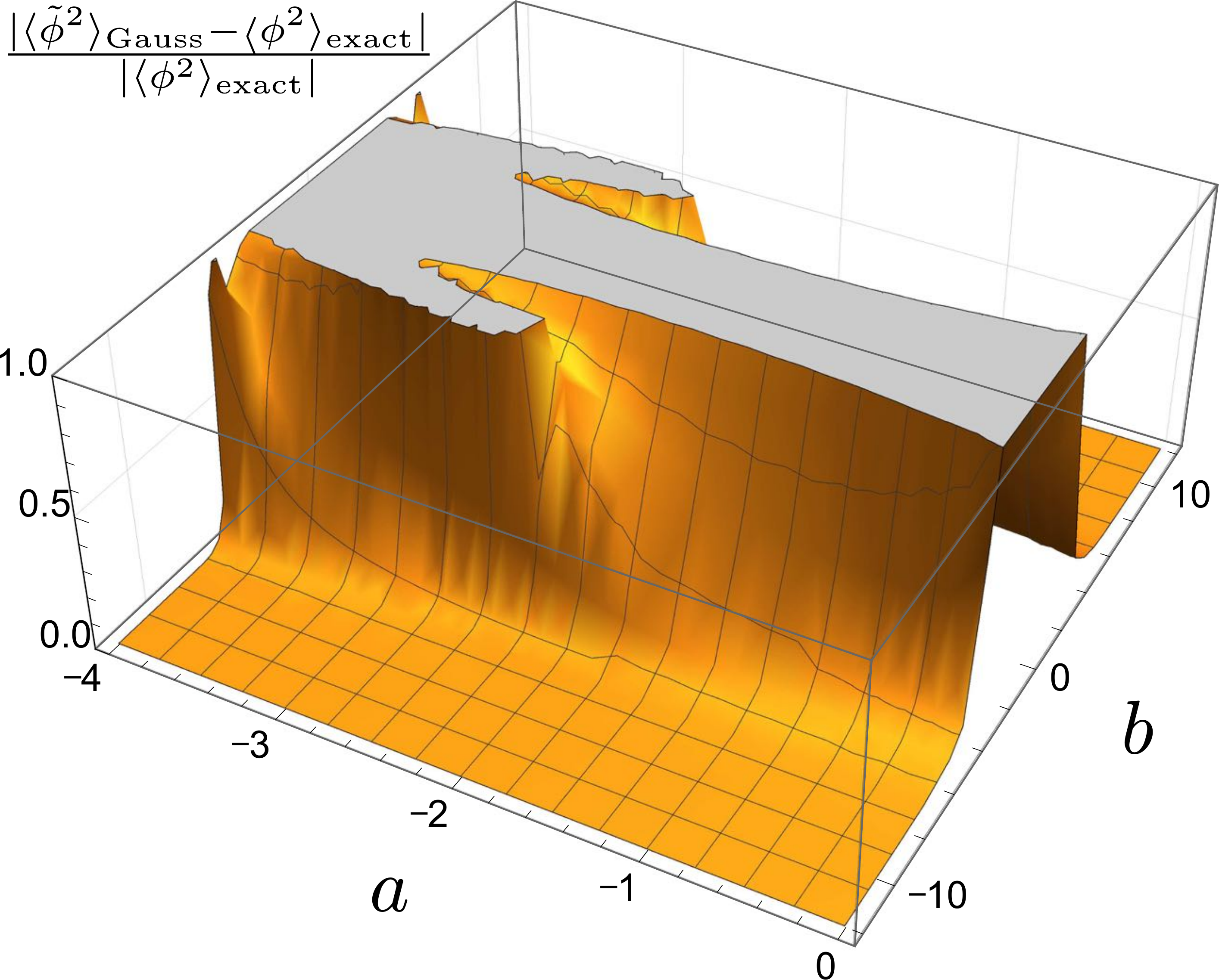}
  \caption{Comparison in the $a<0$ region between the Gaussian Ansatz
    results with the unstable branch of solution and the exact
    answer.}
  \label{fig:tilded}
\end{figure}
%---   figure   ---$

For the moment let us postpone discussions on the stability but simply
adopt the above branch of solution to evaluate new
$\langle\tilde{\phi}^2\rangle_{\text{Gauss}}=1/(\tilde{A}+i\tilde{B})$
in the $a<0$ region.  We show the comparison between
$\langle\tilde{\phi}^2\rangle_{\text{Gauss}}$ and the exact answer in
Fig.~\ref{fig:tilded} and it is clear from this comparison that this
new $\langle\tilde{\phi}^2\rangle_{\text{Gauss}}$ gives a very good
approximation in the region with $a<0$ and $b^2 > a^2$ (if
sufficiently away from the phase boundary; see also the structure in
Fig.~\ref{fig:Bexact}).  The important observation is that we take the
difference between $\langle\tilde{\phi}^2\rangle_{\text{Gauss}}$ and
$\langle\phi^2\rangle_{\text{exact}}$ before computing its absolute
value, and so the good agreement seen in Fig.~\ref{fig:tilded}
includes the information on the complex phase.  This means that, even
though $\tilde{A}<0$ seems to be not allowed for convergence of the
Gaussian integral, the exact answer indicates that this seemingly
unstable $\tilde{A}<0$ is actually the right physical branch of
solution.

\begin{figure}
  \includegraphics[width=0.8\columnwidth]{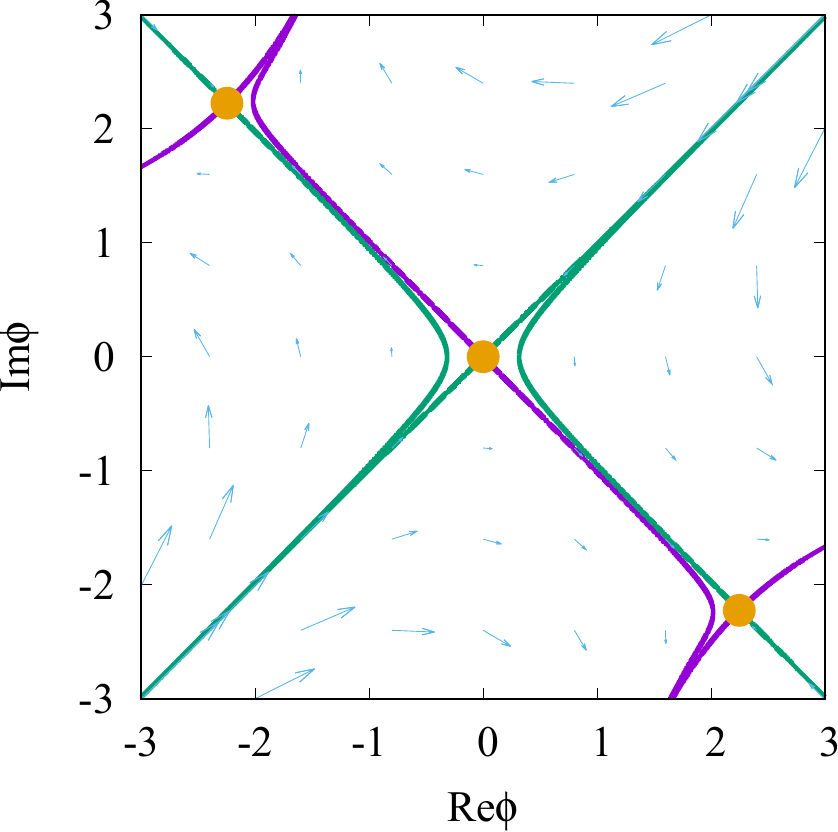}
  \caption{Steepest descendent paths (Lefschetz thimbles) shown by
    purple lines and steepest ascendent paths shown by green lines
    around three saddle-points shown by orange dots.  Parameters are
    chosen as $a=-0.1$ and $b=10$.}
  \label{fig:thimble}
\end{figure}

Then, two questions arise.  One is how such an unstable branch of
solution can be the physical choice.  Another one is what principle
determines which branch of solution describes physically the correct
behavior of the theory.  These questions cannot be answered within the
framework of the CLE or the Gaussian Ansatz of the CLE but we need
more inputs from different approaches.  Here, let us consider these
questions by means of the Lefschetz thimble method.
Figure~\ref{fig:thimble} is a typical example of the thimble
structures in the attenuated state for $a<0$ and $b^2\gg a^2$, and to
draw Fig.~\ref{fig:thimble} we chose $a=-0.1$ and $b=10$.  From this
we understand that the integration path along the thimble or the
steepest descendent path is tilted from the real axis by $-\pi/4$
around $\bar{\phi}_0=0$ if $a<0$ and $b^2\gg a^2$.

%---   figure   ---%
%---   figure   ---$

Because we already know that the integral is dominated by the
contribution near $\bar{\phi}_0$ only, we shall change the integration
variable along the integration path as
\begin{equation}
  \phi\;\longrightarrow\; \phi' = e^{i\pi/4}\phi\;.
\end{equation}
Let us point out that such a U(1) rotation makes sense only in the
complexified description and this kind of transformation may be linked
to the gauge cooling in the gauge theory.  Then, the
action~\eqref{eq:model} near $\bar{\phi}_0$ is expressed as
\begin{equation}
  S \simeq \frac{1}{2}(-i\alpha) \phi'^2
  = \frac{1}{2}(b-ia) \phi'^2\;,
  \label{eq:tilted}
\end{equation}
which implies that the roles of $a$ and $b$ should be switched to each
other along this path.  Needless to say, such a variable change causes
another convergence problem from the $\phi^4$ term in the definition
of the theory, but we are focusing only on the local properties around
$\bar{\phi}_0$.  Such a treatment shall be self-consistently justified
if the probability distribution is localized well around
$\bar{\phi}_0$.  Now, because of Eq.~\eqref{eq:tilted}, the
probability distribution in the Gaussian Ansatz should be parametrized
by $B'$ instead of $A$ and $-A'$ instead of $B$.
Equations~\eqref{eq:MFeq} should be also replaced as
\begin{align}
  B' - b &= \frac{3}{A'^2+B'^2}\cdot (-B')\;,\\
  -A' + a &= \frac{3}{A'^2+B'^2}(-A')\;,
\end{align}
where $c=-1$ (after the change of $\phi\to\phi'$) and $d=0$ are
plugged in.  Then, the above equations are completely identical to the
original equations~\eqref{eq:MFeq} having the same solutions of
Eqs.~\eqref{eq:MFA}, \eqref{eq:MFB}, and Eqs.~\eqref{eq:MFA2},
\eqref{eq:MFB2}.  Even though the solutions are just the same, this
argument tells us an important implication -- for the integration with
the Gaussian Ansatz to converge around $\bar{\phi}_0$, what we need is
$B>0$ and the sign of $A$ does not matter!  In this way, the first
question is answered now.  Both branches of $A$, $B$ and
$\tilde{A}$, $\tilde{B}$ are possible;  Eqs.~\eqref{eq:MFB} and
\eqref{eq:MFB2} lead to $B>0$ and $\tilde{B}>0$ for $b>0$.  We note
that the thimble would be tilted in the opposite direction for $b<0$
and then the convergence would require $B<0$ and $\tilde{B}<0$ which
are also satisfied in Eqs.~\eqref{eq:MFB} and \eqref{eq:MFB2}.

The second question is a little more non-trivial;  both solutions
seem to be equally possible from the integration stability condition.
Besides, in the $a>0$ region, we have chosen
Eqs.~\eqref{eq:MFA} and \eqref{eq:MFB} from the condition of $A>0$ in
the previous discussion, but even in this $A>0$ case, once we consider
the path deformation along the Lefschetz thimble, the convergence
condition around $\bar{\phi}_0$ is only the realness of $A$ and $B$.
Hence, regardless of the sign of $A$, both solutions,
Eqs.~\eqref{eq:MFA}, \eqref{eq:MFB} and Eqs.~\eqref{eq:MFA2},
\eqref{eq:MFB2}, are possible.

%---   figure   ---%
\begin{figure}
  \includegraphics[width=\columnwidth]{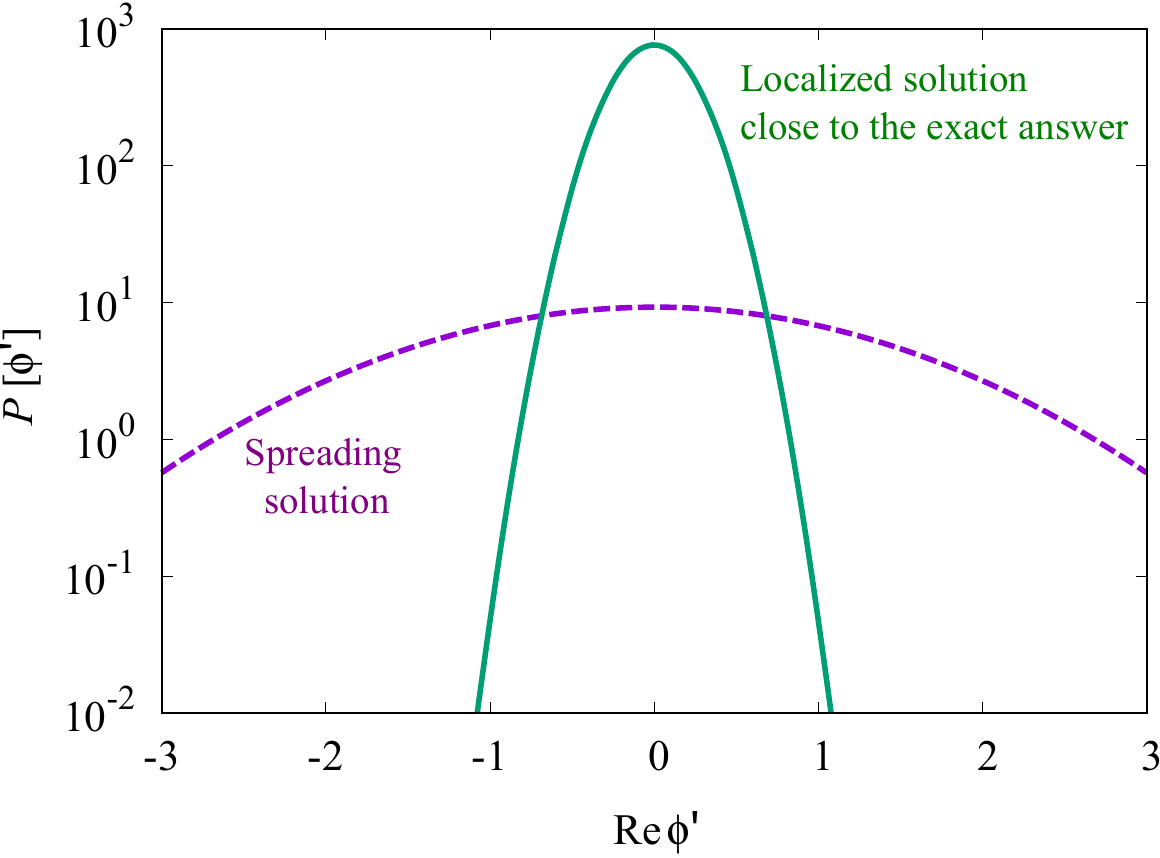}
  \caption{Gaussian profiles along the thimble near $\bar{\phi}_0$
    corresponding to Eqs.~\eqref{eq:MFA}, \eqref{eq:MFB} (spreading
    solution by the dotted line) and Eqs.~\eqref{eq:MFA2},
    \eqref{eq:MFB2} (localized solution by the solid line) for
    $a=-0.1$ and $b=10$.}
  \label{fig:dist}
\end{figure}
%---   figure   ---$

For the self-consistent justification, the Gaussian width must be
small.  The Gaussian width along the thimble near $\bar{\phi}_0$ is
determined by $\sqrt{A^2+B^2}$, and interestingly, we can readily find
from Eqs.~\eqref{eq:MFA}, \eqref{eq:MFB}, and Eqs.~\eqref{eq:MFA2},
\eqref{eq:MFB2} that
\begin{equation}
  \begin{split}
    &\sqrt{A^2+B^2}\gg \sqrt{\tilde{A}^2+\tilde{B}^2}
    \quad \text{ for } \quad a>0\;, \\
    &\sqrt{\tilde{A}^2+\tilde{B}^2}\gg \sqrt{A^2+B^2}
    \quad \text{ for } \quad a<0\;.
  \end{split}
\end{equation}
This observation nicely explains which branch of solution should be
picked up to describe the correct physics.  For the demonstration
purpose to visualize how the probability distributions spread, we plot
$P[\phi']$ along the thimble in terms of $\phi'$ in
Fig.~\ref{fig:dist} and from this plot we see that the spreading
distribution with Eqs.~\eqref{eq:MFA}, \eqref{eq:MFB} is not very well
localized around $\bar{\phi}_0$ but it is stretched to the regions
around $\bar{\phi}_\pm$, which already signals for falling into a
wrong answer.

An interesting question is what then happens if we implement the CLE
simulation after performing a rotation like Eq.~\eqref{eq:tilted} or
more generally: $S[\phi']=\frac{1}{2}|\alpha|\phi'^2
+\frac{1}{4}\beta\,e^{-2i\theta}\phi'^4$ where $\theta$ is an
argument of a complex number $\alpha$ and
$\phi'=e^{i\theta/2}\phi$.  Such a deformation of the theory
might make the existence of the theory questionable, but we should
remember that the existence of the theory is already subtle as soon as
it is complexified because the complexified stochastic processes may
always hit diverging rays on the complex plane;  for the existence of
the theory due to the cancellation of divergences, see discussions in
Sec.~11.5 in Ref.~\cite{Damgaard:1987rr}.  We may also say that we
could have put a small $\phi'^4$ term in the probability distribution
that guarantees the convergence, which is dropped in the Gaussian
Ansatz.

%---   figure   ---%
\begin{figure}
  \includegraphics[width=\columnwidth]{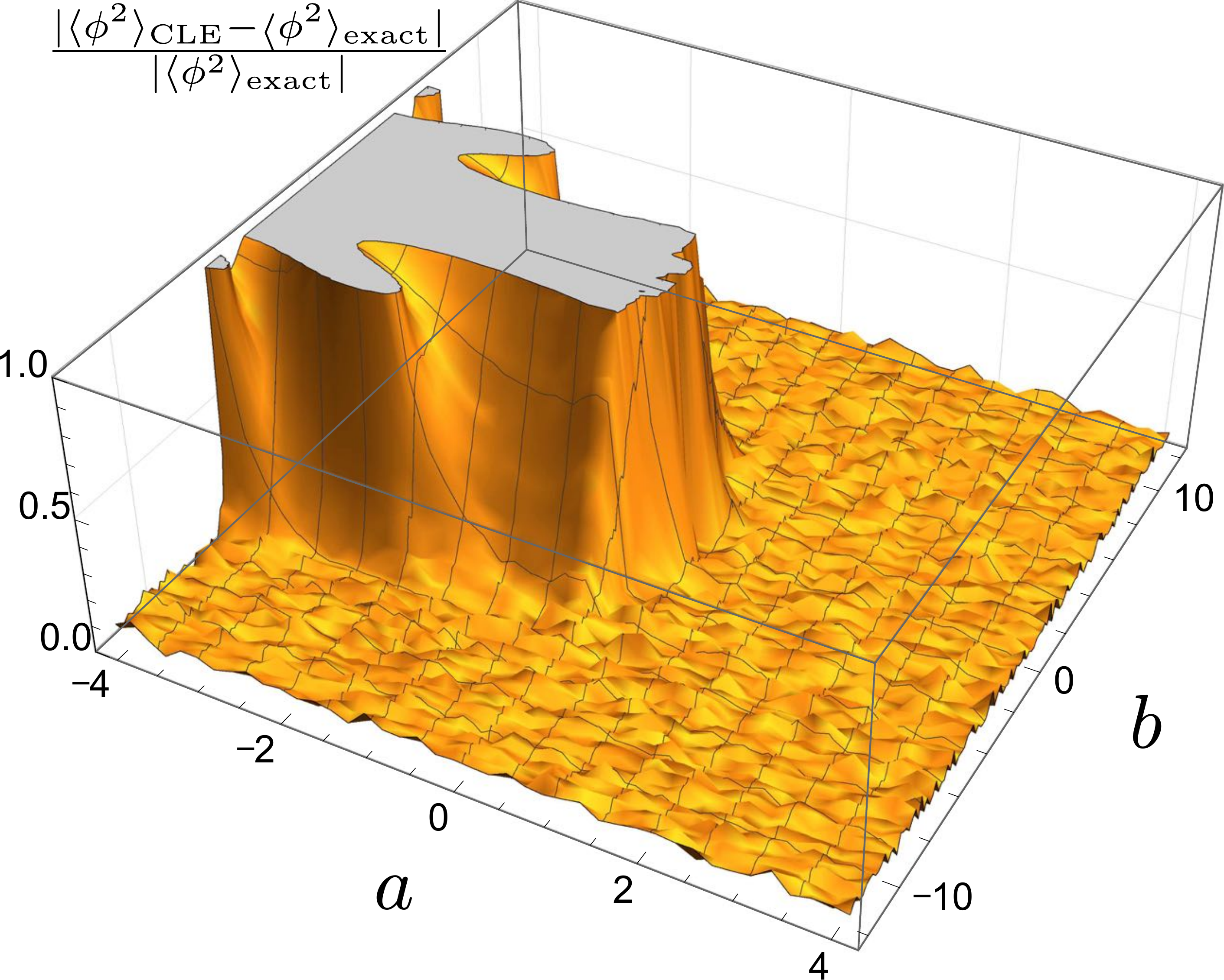}
  \caption{Comparison between the CLE results along the Lefschetz
    thimble and the exact answer for the quadratic operator.}
  \label{fig:cle2}
\end{figure}
%---   figure   ---$

Figure~\ref{fig:cle2} shows our CLE results along the tilted path in
terms of $\phi'$.  The structure for $a<0$ is surprisingly
similar to the results in the Gaussian Ansatz with $\tilde{A}$ and
$\tilde{B}$ presented in Fig.~\ref{fig:tilded}.  This is a very clear
numerical evidence about such a close relation between the CLE and the
Gaussian Ansatz results.  Summarizing our findings, the failure of the
CLE simulation in the attenuated state as seen in Fig.~\ref{fig:cle}
is attributed to the seemingly more stable but unphysically spreading
solution, and if the CLE simulation is performed along the Lefschetz
thimble without phase oscillation, the physical well-localized
solution is correctly picked up to recover the exact answer.  Again,
we would emphasize that our proposed simple treatment of the Gaussian
Ansatz is so powerful to explain all the good and bad behavior of the
CLE simulations.

%---   figure   ---%
\begin{figure}
  \includegraphics[width=\columnwidth]{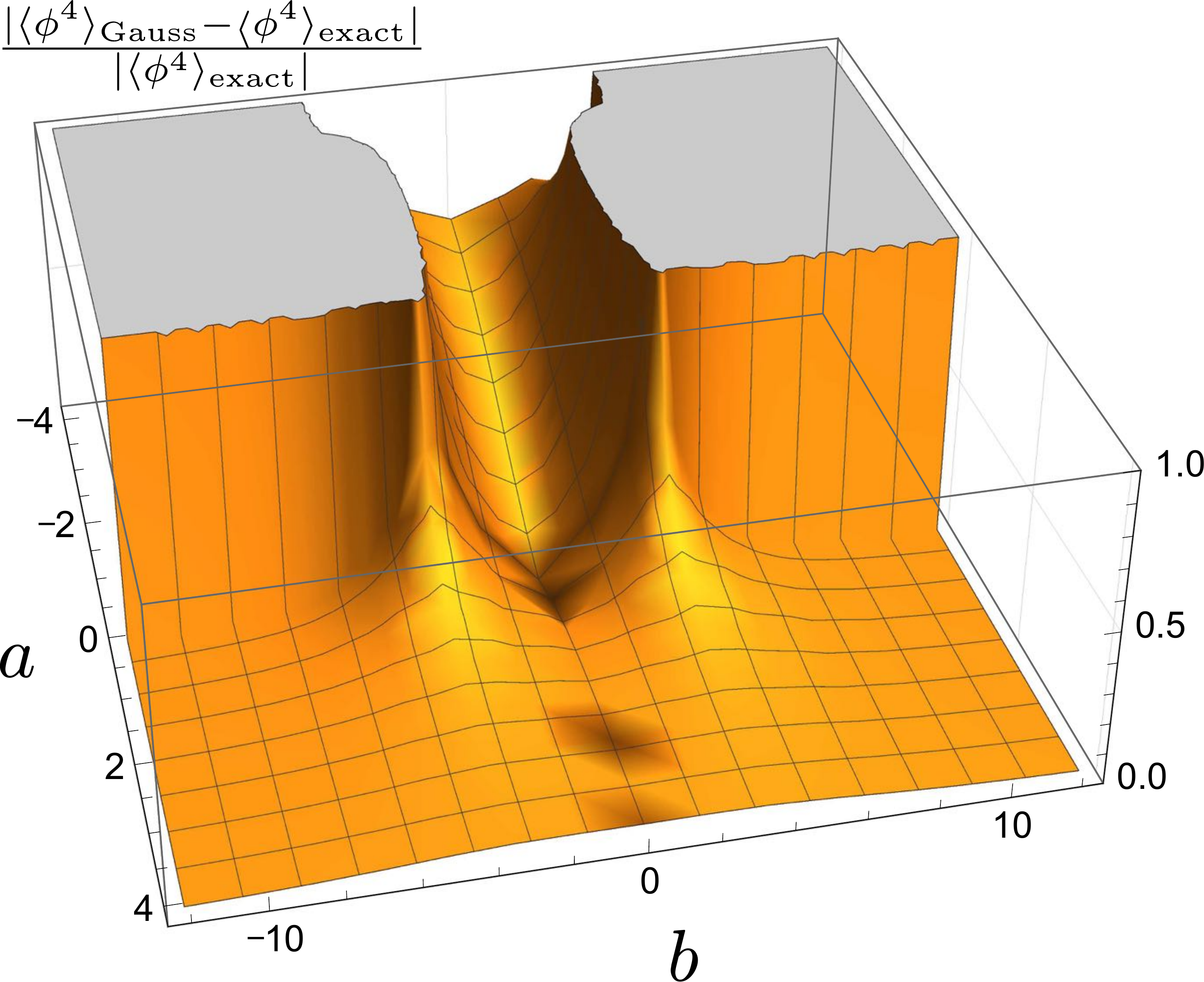}
  \caption{Comparison between the Gaussian Ansatz results and the
    exact answer for the quartic operator.}
  \label{fig:gauss4}
\end{figure}
%---   figure   ---$

Another interesting and important question is whether the Gaussian
Ansatz works for higher order operators or not.  It has been
argued in Ref.~\cite{Aarts:2013uza} that $\langle\phi^4\rangle$ cannot
converge to the physical answer for $b^2>3a^2$ even in the $a>0$
region, while $\langle\phi^2\rangle$ can.  Our optimistic guess is
that the Gaussian Ansatz should be a valid description for higher
order operators because the probability distribution is exponentially
localized by construction.  It is quite easy to check it by a
generalization of
Eq.~\eqref{eq:MF-GF}, i.e.
\begin{equation}
  \langle\phi^4\rangle_{\text{Gauss}}
  = \frac{3}{[A(\alpha,\beta) + i B(\alpha,\beta)]^2}\;.
\end{equation}
Then, the comparison to the exact answer is shown in
Fig.~\ref{fig:gauss4}, which looks just like the comparison for
$\langle\phi^2\rangle$ in Fig.~\ref{fig:gaussNS}.  From this explicit
comparison it is obvious that the Gaussian Ansatz definitely remains
as a good approximation even for higher order operators in the region
with $a>0$ and $b^2>3a^2$.

\begin{figure}
  \includegraphics[width=\columnwidth]{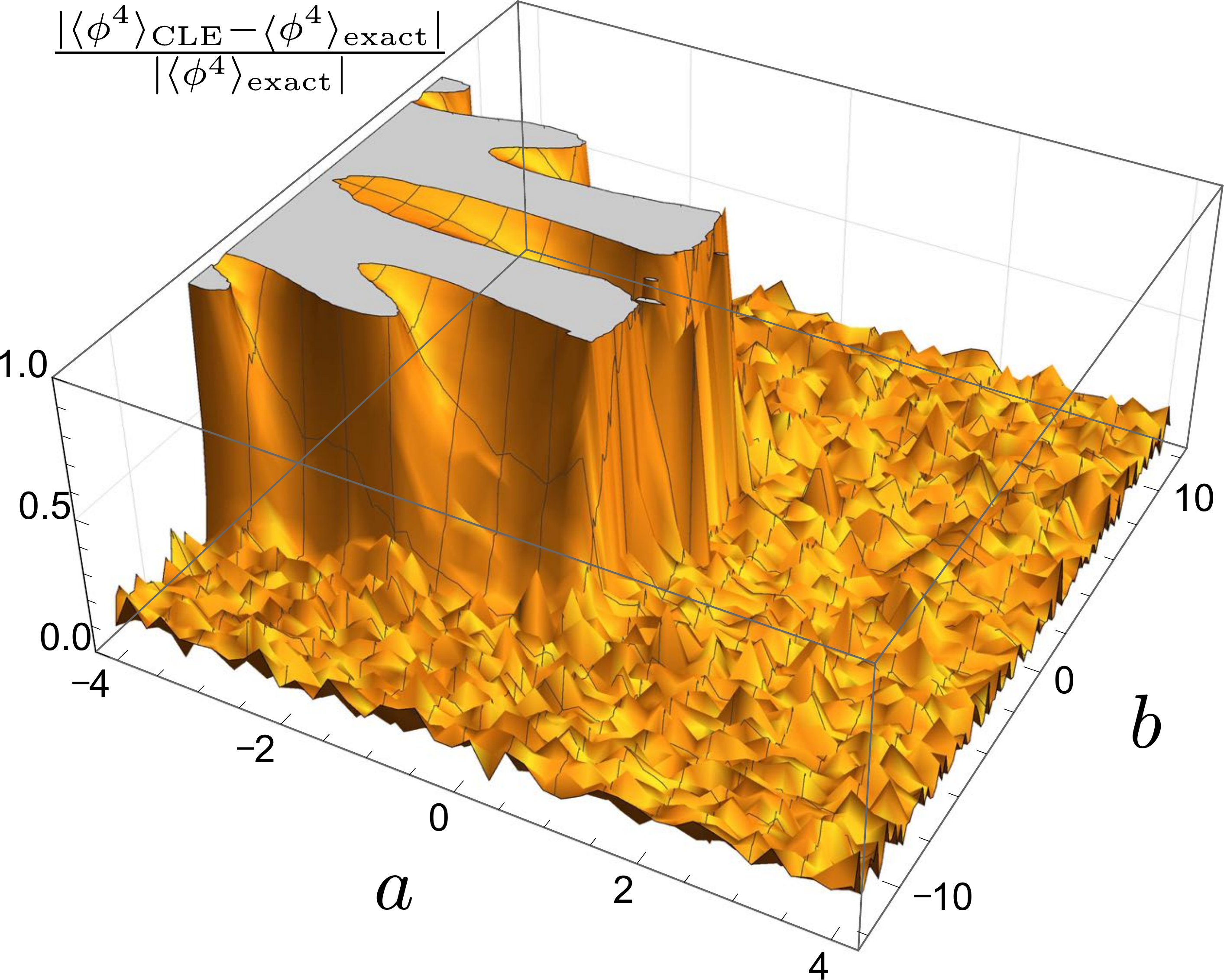}
  \caption{Comparison between the CLE results along the Lefschetz
    thimble and the exact answer for the quartic operator.}
  \label{fig:cle4}
\end{figure}

Now, a natural question is what happens in the numerical CLE
simulation for $\langle\phi^4\rangle$.  If the original action is put
into the CLE simulation, we can indeed see sizable deviations of
$\langle\phi^4\rangle_{\text{CLE}}$ from the exact answer even at
$a>0$ in the $b^2\gg a^2$ region.  If the simulation goes to higher
and higher order operators, the failure region expands toward
$b^2>3a^2$ gradually.  Now, the most interesting question is whether
the transformation from $\phi$ to $\phi'$ can help the CLE simulation
with approaching the exact answer.  Figure~\ref{fig:cle4} shows such a
comparison.  Surprisingly, it is evident that the convergence problem
in the $a>0$ region has been completely resolved.

%---   figure   ---%
%---   figure   ---$

%%%%%%%%%%   Conclusion   %%%%%%%%%%
\section{Conclusion}
\label{sec:conclusion}

We have exploited an approximation scheme using the Gaussian Ansatz to
solve the Fokker-Planck equation and made quantitative comparisons
between the complex Langevin equation, CLE, results and the Gaussian
Ansatz results.  Although the gap equations in the Gaussian Ansatz are
so simple, we have confirmed that multiple solutions from them capture
all the essential features of the CLE simulation not only for the
successful regions but also for the unsuccessful regions.

Our most striking finding is that in the unsuccessful parameter
regions one branch of solutions that is poorly localized is picked up
by the Gaussian Ansatz and the CLE simulation, but another branch of
solutions corresponds to the correct physical answer.  To pick the
correct one up, if the theory is reexpressed in terms of the variable
along the Lefschetz thimble, the branch of solution that describes a
more localized probability distribution can be picked up, which has
been revealed by multiple solution structures in the Gaussian Ansatz.
The idea has been tested in the CLE simulation also, and we have
verified the consistency between the thimble-guided CLE simulation and
the physical branch of solutions in the Gaussian Ansatz.

Our analysis implies a positive prospect in favor of the CLE
simulation.  Even when the convergence property is not sufficiently
good leading to unphysical results, it does not necessarily mean the
complete breakdown of the method itself.  When it occurs, the CLE
simulation may have multiple fixed points and it simply falls into
where it should not fall into, though one of other fixed points may
still correspond to the correct physics.  This is very nice, because a
minimal change in the treatment like the deformation of the
integration path could help us with digging out the correct branch of
solutions (see also the results in
Refs.~\cite{Aarts:2012ft,Tsutsui:2015tua} for other minimal changes to
improve the correct convergence).  This observation might have
something to do with recently proposed ideas on the gauge cooling for
the singular-drift problem~\cite{Nagata:2015uga}.  It would be the
most exciting challenge to apply our method of the Gaussian Ansatz to
another theory like
$S=-(\beta\cos\theta + i\theta)$~\cite{Ambjorn:1985iw} which is the
simplest integral that emulates the symmetry properties of lattice
gauge theories.

\acknowledgments
The authors thank
Gert~Aarts,
Yoshimasa~Hidaka,
Jan~Pawlowski,
and
Yuya~Tanizaki
for useful discussions and comments.
K.~F.\ is grateful for a warm hospitality at
Institut f\"{u}r Theoretische Physik,
Universit\"{a}t Heidelberg, where K.~F.\ stayed as a visiting
professor of EMMI-ExtreMe Matter Institute/GSI and a part of this work
was completed there.
This work was partially supported by JSPS KAKENHI Grant
No.\ 15H03652 and 15K13479.

\end{document}